\documentclass[12pt]{article}
\usepackage{amssymb}
\usepackage{dsfont}
\normalsize
\input{epsf}

\begin{document}
\addtolength{\baselineskip}{.20mm}
\newlength{\extraspace}
\setlength{\extraspace}{2mm}
\newlength{\extraspaces}
\setlength{\extraspaces}{2mm}

\newcommand{\newsection}[1]{
\vspace{15mm} \pagebreak[3] \addtocounter{section}{1}
\setcounter{subsection}{0} \setcounter{footnote}{0}
\noindent {\Large\bf \thesection. #1} \nopagebreak
\medskip
\nopagebreak}
\newcommand{\newsubsection}[1]{
\vspace{1cm} \pagebreak[3] \addtocounter{subsection}{1}
\addcontentsline{toc}{subsection}{\protect
\numberline{\arabic{section}.\arabic{subsection}}{#1}}
\noindent{\large\bf 
\thesubsection. #1} \nopagebreak \vspace{3mm} \nopagebreak}
\newcommand{\ba}{\begin{eqnarray}
\addtolength{\abovedisplayskip}{\extraspaces}
\addtolength{\belowdisplayskip}{\extraspaces}

\addtolength{\belowdisplayshortskip}{\extraspace}}

\newcommand{\be}{\begin{equation}
\addtolength{\abovedisplayskip}{\extraspaces}
\addtolength{\belowdisplayskip}{\extraspaces}
\addtolength{\abovedisplayshortskip}{\extraspace}
\addtolength{\belowdisplayshortskip}{\extraspace}}
\newcommand{\ee}{\end{equation}}
\newcommand{\STr}{{\rm STr}}
\newcommand{\figuur}[3]{
\begin{figure}[t]\begin{center}
\leavevmode\hbox{\epsfxsize=#2 \epsffile{#1.eps}}\\[3mm]
\parbox{15.5cm}{\small
\it #3}
\end{center}
\end{figure}}
\newcommand{\im}{{\rm Im}}
\newcommand{\calm}{{\cal M}}
\newcommand{\call}{{\cal L}}
\newcommand{\sect}[1]{\section{#1}}
\newcommand\hi{{\rm i}}
\def\bea{\begin{eqnarray}}
\def\eea{\end{eqnarray}}

\begin{titlepage}
\begin{center}

\vspace{3.5cm}

{\Large \bf{The  Unified First law  in ``Cosmic Triad" Vector Field Scenario}}\\[1.5cm]

{Yi Zhang $^{a,b,}$\footnote{Email: zhangyia@cqupt.edu.cn},}{Yungui
Gong $^{a,}$\footnote{Email: gongyg@cqupt.edu.cn}, }{Zong-Hong Zhu
$^{b,}$\footnote{Email: zhuzh@bnu.edu.cn},} \vspace*{0.5cm}

{\it $^{a}$College of Mathematics and Physics, Chongqing University of
Posts and Telecommunications, \\ Chongqing 400065, China

 $^{b}$ Department of Astronomy, Beijing Normal university,
\\  Beijing 100875, China}

\date{\today}
\vspace{3.5cm}

\textbf{Abstract} \vspace{5mm}

\end{center}
 In this letter, we try to apply the unified first law to the ``cosmic
triad" vector field scenario both in  the minimal coupling case and in the non-minimal
coupling case. After transferring the non-minimally coupling action
 in  Jordan frame
 to  Einstein frame,   the correct dynamical equation (Friedmann
 equation) is gotten in a thermal equilibrium process  by using the already existing
 entropy while the entropy in the non-minimal coupled ``cosmic triad" scenario has not
 been derived.
 And after transferring the variables back to Jordan frame, the corresponding Friedmann equation is  demonstrated to be correct.
For  complete arguments,  we also calculate
the related  Misner-Sharp energy in  Jordan and Einstein frames.

\end{titlepage}

\section{Introduction}\label{sec1}

 The profound connections between gravity and
thermodynamics  are suggested by many phenomena, such as the
discovery of Hawking
 radiation  and the
 four  laws of classical black hole mechanics \cite{Bardeen:1973gs,Bekenstein:1973ur,Hawking:1974sw,Gibbons:1977mu}.
  Based on the geometric feature of thermodynamic
quantities of black holes \cite{Jacobson:1995ab}, a remarkable connection for cosmology
is found by  Jacobson  who pointed out it is possible to derive the
Einstein equations of gravitational fields from a  view  of
thermodynamics. The keys to derive Einstein
equation are  the fundamental relation (Clausius relation) $\delta
Q=Td S$ and  the form of the entropy which is  proportional to the horizon area. Further studies between  thermodynamics and  gravity
have been extended to various cosmological settings
\cite{Padmanabhan:2002sha,Frolov:2002va,Danielsson:2004xw,Bousso:2004tv,Cai:2005ra,Akbar:2008vc},
such as the Lovelock gravity
\cite{Akbar:2006er,Akbar:2006kj,TD-LL-Pad,Cai:2008mh,Cai:2009de},
the brane-world scenario
\cite{Braneworld-scenario,Gong:2007md,Sakai:2001gh},   the
scalar-tensor theory
\cite{Gong:2006ma,Gong:2006sn,Saridakis:2009uu,Wang:2005bi}, the
loop quantum gravity \cite{loopKA,positive,loop,Cai:2008ys,LQC-Zhu},
the Horava-Lifshitz
 gravity \cite{HL-gravity},
 the
logarithm correction scenario
\cite{32,12,0,flu,Zhu:2008cg,Cai:2009ua},  the trace anomaly
correction scenario \cite{Lidsey:2008zq},  and the $f(R)$ gravity
\cite{Akbar:2006mq,Cai:2006rs,Chirco:2009dc,Wu:2007se,Elizalde:2008pv,Wald
entropy,Brustein:2009hy,Bamba:2009id,Bamba:2009gq,Bamba:2010kf,Parikh:2009qs,Brustein:2007jj,Jacobson:1993vj,Cognola:2005de}.

In cosmology, the scalar field  could be assumed to be isotropic and
homogeneous to correspond with the FRW ( Friedmann-Robertson-Walker) background. It is the most
popular candidate of  dynamical sources for the  accelerations in
our universe
\cite{Guth:1980zm,Linde:1981mu,Riess:1998cb,Perlmutter:1998np}.
However, the fundamental scalar
 field \cite{Kofman:2007tr,Linde:2001ae} has not  to be probed yet. On the other hand,
 the vector field is common in our realistic world. The  inflationary scenario  with vector
fields was proposed by
Refs.\cite{Ford:1988wq,Ford:1989me}.  Despite  the later discovered  instability problems \cite{Himmetoglu:2008zp} in  perturbations \cite{Dimopoulos:2006ms,Dimopoulos:2008rf,Dimopoulos:2008yv}, this vector field scenario was even extended to higher spin
fields  in cosmology \cite{Germani:2009iq,Kobayashi:2009hj,Koivisto:2009sd}.  The
``cosmic triad" scenario is one of those models that coincide with the observable
isotropic and homogenous FRW background
\cite{Bento:1992wy,ArmendarizPicon:2004pm,Zhang:2009yu,Zhang:2009tj,Wei:2006tn,Chiba:2008rp}£¬(see
also `N-inflation'' vector field scenario proposed by
Refs.\cite{Golovnev:2008cf,Golovnev:2008hv,Golovnev:2009ks} which is similar to ``N-flation" in scalar field \cite{Dimopoulos:2005ac},
the time-like vector filed scenario proposed by Refs.\cite{Kiselev:2004py,Carroll:2004ai, Lim:2004js,Boehmer:2007qa,
Koivisto:2008xf,Koh:2009vm,Koh:2009ne,Carroll:2009em,Li:2007vz}, and the exact isotropic solutions of the Einstein-Yang-Mills system
 proposed by
Refs.\cite{Hosotani:1984wj,Galtsov:1991un,Zhao:2005bu,Zhao:2006mk}). The  ``cosmic triad" scenario of vector field has  three spatial components
 equal  and orthogonal to each other where $A_{1}=(0,A,0,0)$, $A_{2}=(0,0,A,0)$ and $A_{3}=(0,0,0,A)$.
 In this letter, we will use view  to
study the relation between  gravity and  thermodynamics.

In  a special kind of spherically symmetric black hole
space-times, Padmanabhan showed that the Einstein equations on the
back hole could be written into  the first law of thermodynamics
\cite{Padmanabhan:2002sha}. Cai and Kim \cite{Cai:2005ra} derived
the Friedmann equation by assuming
that the apparent horizon has temperature and entropy  and applying the fundamental relation $\delta
Q=T \delta S$ to the apparent horizon of FRW universe. The
Clausius relation requires the equilibrium
  thermodynamics.
In Einstein gravity, the Clausius relation for the equilibrium
thermodynamics  could always hold. However, there are some
arguments on  the existence of
thermal equilibrium process for the non-Einstein gravity, such as the scalar-tensor theory
( the $f(R)$ theory as well). The
field equation for scalar-tensor gravity needs the non-equilibrium
thermodynamics arguments in Refs.\cite{Eling:2006aw,Eling:2008af}.
To get the Friedmann equation, the related thermal dynamical
discussion has used the bulk viscosity entropy production term
\cite{Akbar:2006mq,Cai:2006rs,Chirco:2009dc,Wu:2007se}. Therefore,
it is   proposed to add the entropy production term to get the
Friedmann equation in Ref.\cite{Cai:2006rs}.  Meanwhile, it
was noticed that the entropy of static horizon is well defined by
Wald's definition in Refs.\cite{Elizalde:2008pv,Wu:2009wp,Briscese:2007cd}, which is a Noether charge associated with the horizon
killing vector and  the correct Friedmann equation could be gotten
in the non-minimally coupled gravity with equilibrium thermodynamics.

The non-minimally coupled vector fields  bring us a new physical
background
\cite{Golovnev:2008cf,Golovnev:2008hv,Golovnev:2009ks}. The
non-minimally coupled ``cosmic triad" vector field scenario is quite
similar to the scalar-tensor theories of gravity. Therefore, it is rather natural to ask
whether the corresponding physical process is thermal
equilibrium or not. Even worse, we have no idea of  the entropy definition in the non-minimally coupled ``cosmic
triad" vector field scenario. Fortunately,  the Einstein frame could be used
as a bridge. The exact form of  the entropy of  ``cosmic triad" scenario in the Jordan frame is
not prerequisite. Based on the conformal transformations and the
entropy form of the Einstein gravity, we can still derive the Friedmann equation.

Our derivations of the
Friedmann equation will also be affected by the definition of  energy.
 To make our arguments complete and consistent, we try to discuss general Misner-Sharp energies
\cite{Misner:1964je,Bak:1999hd} in a spherically symmetric spacetime
 by  integral method. The generalized Misner-Sharp energy is argued to
be related to the Einstein equation whose definition is clear in the Einstein
gravity, but not in the non-Einstein gravity \cite{Misner:1964je,Bak:1999hd}. The thermal
equilibrium process in scalar-tensor gravity has been presented in Ref.
\cite{Akbar:2006mq,Cai:2006rs,Chirco:2009dc,Wu:2007se}  the effective geometric part included in the total energy density.
However, our results will not include  the obvious effective geometric part.
 We use
units of $k_{B}=c=\hbar=1$ and denote the gravitational constant
$8\pi G $ by $\kappa^{2}=8\pi m_{pl}^{-2}$ where $m_{pl}=G^{-1/2}$ is
the Planck mass.

We arrange our letter as follows. In Sec.\ref{sec2}, we introduce basic
notions in thermodynamics,   the temperature, the apparent horizon, the unified first
law and the Clausius relation. After that, we present the minimally coupled ``cosmic triad" vector field
model and deduce its dynamical equation in Sec.\ref{sec3}. Then,  in the non-minimal coupling case, considering the
similarity between  scalar-tensor theories and the discussed vector fields theory,  we manage to
get the Friedmann equation  with the
help of  Einstein frame in Sec.\ref{sec4}. For consistency,   the
results of the general Misner-Sharp energy are presented by  integral
method in Sec.\ref{sec5}. The paper is concluded in
Sec.\ref{sec6}.

 \section{The Unified First Law }\label{sec2}

 The FRW metric is one kind of
 spherically symmetric space time. If the closure of a
 hypersurface was foliated by future or past, outer or
 inner marginal sphere, it  is the so-called trapping horizon.  However, in the
FRW universe, the ``outer trapping horizon" does not exist, instead
there are a kind of cosmological horizons called ``inner trapping
horizon" which is  the apparent horizon in the context of the FRW
cosmology. In this letter, we will not distinguish the two horizons.
And, the associated thermodynamics will be discussed. The
$(3+1)$-dimensional FRW universe has the metric
 \be
 \label{frwmetric}
 ds^{2}=-dt^{2}+a^{2}\gamma_{ij}dx^{i}dx^{j},
 \ee
 where $a$ is the scale factor,
 the metric $\gamma_{ij}$ is given by $
 \gamma_{ij}=d\rho^{2}/(1-k\rho^{2})+\rho^{2}d\Omega_{n-1}^{2}$  and the three-dimensional spatial  curvature of the hypersurface is parameterized as negative, zero or
 positive, respectively. The
FRW metric could be rewritten in the double null form as well
 \be
 ds^{2}=h_{ab}dx^{a}dx^{b}+r^{2}d\Omega_{n-1}^{2},
 \ee
 where $r=a(t)\rho$, $x^{0}=t$, $x^{1}=\rho$ and the two
 dimensional metric is $h_{ab}=diag(-1,1/(1-k\rho^{2}))$.

The thermodynamics will be established on the apparent horizon where
 the future inner trapping horizon is the boundary of
 a
 system. The dynamical
 apparent horizon is defined as
  \be
  r_{A}=h_{ab}\partial_{a}r\partial_{b}r=\frac{1}{\sqrt{H^{2}+k/a^{2}}},
  \ee
   where $H=\dot{a}/a$ is the Hubble parameter.
And, the surface gravity of the trapping horizon $\kappa_{s}$ is
defined as
 \be
 \kappa_{s}=\frac{1}{2}\nabla^{a}\nabla_{a}r=r_{A}(1-\frac{\dot{r}_{A}}{2Hr_{A}}),
 \ee
 where the subcript ``$s$'' is used to note the variables for the
 thermodynamics specially. Then, the corresponding temperature is
 \be
 \label{T}
 T=\frac{\kappa_{s}}{2\pi}=-\frac{r_{A}}{2}(\dot{H}+2H^{2}+\frac{k}{a^{2}}).
  \ee

For dynamical black holes, Hayward \cite{Hayward2,Hayward3} has
proposed a relation called ``unified first law" to deal with the
gravity and the thermodynamics associated with trapping horizon of a
dynamical black hole in four-dimensional Einstein theory. For
spherically symmetrical space-times, the time-time component of the
Einstein equations could be rewritten in  the ``unified first law"
form
 \begin{equation}
 \label{uni}
 d E = A_{s} \Psi + W d V,
 \end{equation}
 where $A_{s} $ and  $V$ are the area and volume of the three-dimensional
  space. The first term in the unified first law could be interpreted as
an energy-supply term, analogous  to the heat-supply term in the
classical first law of thermodynamics.
 One has
  \be
   \label{Psi}
\Psi=\Psi_{t}dt+\Psi_{\rho}d\rho=-\frac{1}{2}(\rho^{(m)}+p^{(m)})H r
dt+\frac{1}{2}(\rho^{(m)}+p^{(m)})a d\rho,
  \ee
  where the superscript ``$(m)$" notes the variables for  the total matter which includes  not only the
 pressure matter, but also the matter field part. In this letter,  we just neglect the radiation part which
 could be added conveniently by rewriting the Lagrangian and it  would not affect our results.
And, the second term in the unified first law could  be interpreted as
a work term. Following Refs.
\cite{Cai:2005ra,Bak:1999hd,Hayward2,Hayward3},
 the work density at the apparent horizon is
\begin{equation}
\label{W}
 W = -\frac{1}{2}T^{ab}h_{ab}=\frac{1}{2}(\rho^{(m)}-p^{(m)}),
 \end{equation}
which should be regarded as
 the work done by a change of the apparent horizon. Finally, on the left hand side
 of the unified first law,
the energy on the apparent horizon is the generalized Misner-Sharp
energy
  \be
  \label{EV}
 dE=A_{s}\Psi+A_{s}Wd r_{A}=-(\rho^{(m)}+p^{(m)})A_{s}H r_{A}dt +A_{s}\rho^{(m)} dr_{A}=d(\rho^{(m)}
 V).
  \ee

On the other side, during the time interval $dt$, the Clausius relation gives out an
energy flux
  \be
 \label{clausius}
 \delta Q= TdS ,
  \ee
 where $\delta Q$ and $T$ are the variation of heat
flow and the Unruh temperature seen by an accelerated observer just
inside the horizon. Then,  by matching the heat flux of energy and
the amount of energy crossing the apparent horizon, one has
 \be
 \label{deltaQ}
  \delta Q=TdS=A\Psi.
   \ee
In
 Einstein gravity, the unified first law also implies the Clausius
 relation $\delta Q=TdS$ \cite{Cai:2009qf}. The Clausius relation holds for all local Rindler causal
horizon through each spacetime point in the equilibrium
thermodynamics.
 Therefore, in  equilibrium thermodynamics, by matching Eqs.(\ref{uni}) and (\ref{clausius}), it obtains
  \be
  \label{crit}
  TdS=d E-WdV.
  \ee
  Combined with the temperature (\ref{T}), the above
  equation could be rewritten as
 \be \label{ess}
\frac{1}{2\pi
 r_{A}}(1-\frac{\dot{r}_{A}}{2Hr_{A}})dS=4\pi r^{3}_{A}H(\rho^{(m)}+p^{(m)})dt- 2\pi
r^{2}_{A}(\rho^{(m)}+p^{(m)})dr_{A}
 \ee
where  the equilibrium thermodynamics must hold. It has
been shown  that  the above equation is held in  Einstein
gravity with  the pressureless matter. However, if
 the vector fields were added, it is a question  whether this
situation will be changed or not. However,  given the exact form of
 entropy, Eq.(\ref{ess}) will give out the Raychaudhuri equation
which  connects the  geometry
 and  the matter. Furthermore,  by considering the conserved equation
 of the energy density,  the Friedmann equation will be  easily derived.

\section{ ``Cosmic Triad" Vector Field Scenario}\label{sec3}

The ``cosmic triad" vector field scenario
\cite{ArmendarizPicon:2004pm}  is composed of  three vector
fields minimally coupled with gravity, which are a set of three
identical self-interacting vectors. This kind of vector fields
naturally arise  from a gauge theory with $SU(2)$ or $SO(3)$ gauge
group. In this letter, Latin
indices are used to label the different fields ($a,b,...=1...3$), and Greek
indices are used to label the different space-time components
($\mu,\nu,...=0...3$).  In minimal coupling case, the action of  ``cosmic triad"
scenario is
  \be
  \label{action1}
  \mathcal{I}=\int d^{4}x \sqrt{-g}\left[\frac{R}{16 \pi G}-\sum^{3}_{a=1}\left(\frac{1}{4}F^{a}_{\mu\nu}F^{a\mu\nu}+V(A^{a2})\right)+\mathcal{L}_{m}\right],
  \ee
where
$F^{a}_{\mu\nu}=\partial_{\mu}A_{\nu}^{a}-\partial_{\nu}A_{\mu}^{a}$,
$A^{a2}=g^{\mu\nu}A_{\mu}^{a}A_{\nu}^{a}$, $A_{\nu}^{a}$ is the vector field and $\mathcal{L}_{m}$ is the
Lagrangian of  pressureless  matter. The term $ F^{a}_{\mu\nu}F^{a\mu\nu}/4$ could
be considered as the Maxwell type kinetic energy term, and the
term $V(A^{2})$ as the potential of the vector field.  We assume
the energy density of  pressureless matter  conservation
 \be
 \label{conm}
 \dot{\rho}_{m}+3H(\rho_{m}+p_{m})=0,
 \ee
 where the dot means a derivative with respect to time $t$.

In ``cosmic triad" vector field scenario, the ansatz that the three
vectors are equal and orthogonal to each other could be expressed as
  \be
  \label{B}
  A^{b}_{\mu}=\delta^{b}_{\mu}B(t)\cdot a.
 \ee
Following Ref.\cite{ArmendarizPicon:2004pm}, we could define  a
new variable called ``physical" vector field which is
$B_{i}=A_{i}/a$,
 where $ A_{i}$ is called ``comoving" vector field.
The related equation $B^{2}=B_{i}B_{i}=A_{\mu}A^{\mu}=A^{2}$ could
be  conveniently obtained in the FRW background. Then we could express
most of our equations in term of $B_{i}$ and $B^{2}$ in the
following discussions.
  The corresponding
energy density  and pressure are given by
\begin{eqnarray}
\label{rhom}
 &&\rho_{v}=\frac{3}{2}(\dot{B}+H B)^{2}+3V(B^{2}),\\
 \label{pm}
 &&p_{v}=\frac{1}{2}(\dot{B}+H B)^{2}-3V(B^{2})+2B^{2}V',
  \end{eqnarray}
where the subscription ``$v$" means the variable corresponding to the
vector fields,
and the prime  denotes a derivative with respect to the square of  vector field
$B$, for example $V'=dV/dB^{2}=dV/dA^{2}$. In the minimal coupling
case, the energy density of the vector field is conserved as well
 \be
 \label{conv}
 \dot{\rho}_{v}+3H(\rho_{v}+p_{v})=0.
  \ee
  And the equation of  motion of  vector field is
  \be
  \label{eom}
\ddot{B}+3H\dot{B}+
 V'+(2H^{2}+\dot{H})B=0.
  \ee
Obviously, compared with scalar fields, the term $(2H^{2}+\dot{H})B$
is an additional term and therefore the  dynamics of vector field    is
different.

In thermodynamics, there are different definitions of entropy.  Hayward \cite{Hayward2,Hayward3}
has studied blackhole's entropy in generalized theories of gravity
and proposed that the correct dynamical entropy of stationary
blackhole's solution with bifurcate Killing horizon  is  the
Noether charge entropy.  In Einstein gravity, the two definitions
seems to be consistent, the entropy has such a form
 \be
 \label{S1}
 S=\frac{A_{s}}{4\pi G}.
 \ee

Putting the variables  (\ref{rhom}), (\ref{pm}) and (\ref{S1})
 into Eq.(\ref{ess}), we could get
the Raychaudhuri equation in the ``cosmic triad" vector field
scenario
 \be
 \dot{H}-\frac{k}{a^{2}}=-4\pi G\left(2(\dot{B}+H B)^{2}+2B^{2}V'+\rho_{m}+p_{m}\right). \label{frw2}
 \ee

By using the conserved equations (\ref{conm}) and (\ref{conv}), the
Friedmann equation is obtained
 \be
  \label{frw1}
 H^{2}+\frac{k}{a^{2}}=\frac{8\pi G }{3}\left(\frac{3}{2}(\dot{B}+H
 B)^{2}+3V(B^{2})+\rho_{m}\right).
 \ee
During the process, an integral  constant has emerged which
could be  regarded as a cosmological constant and could be incorporated
into the energy density  as a special component. Here, the two
energy components are conserved separately, but
 in the non-minimal coupling case, the situation is more complicated.

\section{Non-Minimally Coupled ``Cosmic Triad" Vector Field Case}\label{sec4}

In the vector field scenario, the non-minimal coupling term is used
to satisfy the slow-roll conditions \cite{Zhang:2009tj}. Without the non-minimal
coupling term, the vector field could only be used as curvaton
\cite{Koivisto:2009sd,Dimopoulos:2006ms,Dimopoulos:2008rf,Dimopoulos:2008yv}.
 Let us start with the  action of  non-minimal coupled ``cosmic triad" vector field
 \be
 \label{actionnon}
\mathcal{I}_{n}=\int d^{4}x \sqrt{-g}\left[\frac{f(A^{2}) R}{16 \pi
G}+3\left(\frac{1}{4}F_{\mu\nu}F^{\mu\nu}-V(A^{2})\right)+\mathcal{L}_{m}(g_{\mu\nu})\right],
 \ee
 where the subscript ``$n$" denotes the non-minimal coupling case,  the function $f(A^{2})$  shows
 the non-minimal coupling effect and it will go back to the minimal coupling case when $f(A^{2})=1$.
Some gauge-dependent second order derivatives of the vector field
$A_{\mu}$ come from the $f(A^{2})R$ term which breaks the gauge
invariance of the vector's kinetic term.

 Conformal (or
Weyl ) transformations are widely used in scalar-tensor theories of
gravity, the theory of  scalar fields coupled nonminimally to the
Ricci curvature $R$. Due to the similarities between   the
scalar-tensor theory and the non-minimal coupling ''vector filed"
scenario, we could perform the conformal transformation from the Jordan
frame to the Einstein frame. One could introduce auxiliary fields,  or even
simply redefine fields for one's convenience. There is no unique
prescription of redefining the fields of a theory. Acting on the
metric by a suitable conformal transformation, the action
(\ref{actionnon}) could be recast into the  one in  Einstein frame
with the new metric,
 \be
 \bar{g}_{\mu\nu}=f(A^{2})g_{\mu\nu},
  \ee
 where the bar represents variables in  Einstein frame. And this frame  is  expected to excite the generic helicity-0 ghost of the
non-invariant vector theories. The corresponding  action in
 Einstein frame is changed to \cite{EspositoFarese:2009aj}
 \be
\bar{I}_{n}=\int d^{4}x \sqrt{\bar{g}}\left[
  \frac{\bar{R}}{2\kappa}-\frac{3}{4\kappa}Z^{2}(\partial_{\mu}\bar{A}^{2})^{2}
  -\frac{1}{4} \bar{F}^{\mu\nu} \bar{F}_{\mu\nu}-W(\bar{A}^{2})\right]+\int d^{4}xL_{m}(\bar{g}_{\mu\nu})
 \ee
where the kinetic terms of the vector $A_{\mu}$ and the tensor
$g_{\mu\nu}$ are now diagonalized in a covariant way, and
 \begin{eqnarray}
&&\bar{F}_{\mu\nu}=F_{\mu\nu}=\partial_{\mu}A_{\nu}-\partial_{\nu}A_{\mu},\,\,\,\,
 \bar{F}^{\mu\nu}=\bar{g}^{\mu\rho}\bar{g}^{\nu\sigma}F_{\rho\sigma}=\partial_{\mu}A_{\nu}-\partial_{\nu}A_{\mu},\\
 &&U=\frac{V}{f^{2}},\,\,\,\,\,\bar{A}^{2}=\frac{A^{2}}{f},\,\,\,\,\,\,Z=-\frac{d ln(1/f)}{d \bar{A}^{2}}=\frac{ff'}{f-A^{2}f'}.
 \end{eqnarray}

The energy densities of  pressureless matter and vector fields are  being rescaled as
 \begin{eqnarray}
&&\bar{\rho}_{m}=\frac{\rho_{m}}{f^{2}},\\
 &&\bar{\rho}_{v}=\frac{3}{2f^{2}}(\dot{B}+H
 B)^{2}+3\frac{V}{f^{2}}+\frac{\dot{f}^{2}}{f^{2}}.
  \end{eqnarray}
  The energy densities of two components are not conserved  separately any more.
  However, the total energy of matter is still
  conserved which includes the rescaled pressureless matter and the rescaled vector fields
  \be
  \label{const}
 \dot{ \bar{\rho}}^{(m)}+3H(\bar{\rho}^{(m)}+\bar{p}^{(m)})=0.
  \ee
  where $\bar{\rho}^{(m)}=\bar{\rho}_{m}+\bar{\rho}_{v}$ and
  $\bar{p}^{(m)}=\bar{p}_{m}+\bar{p}_{v}$.

In  Einstein frame, the entropy could be written as
 \be
 \bar{S}=\frac{\bar{A}_{s}}{4\pi G}.
  \ee
  In order to get the heat $\delta Q$  in the Clausius relation, we have to consider the
contribution from  matter fields. In the Einstein frame, by using
the unified first law,
 one could  get the Raychaudhuri equation
  \be
  \label{bardotH}
  \dot{\bar{H}}-\frac{k}{\bar{a}^{2}}=-4\pi G
  (\bar{\rho}_{v}+\bar{p}_{v}+\bar{\rho}_{m}+\bar{p}_{m}).
  \ee
Combining the above equation with the conserved equations
(\ref{const}),  the Friedmann equation is obtained
  \be
  \label{barh2}
 \bar{H}^{2}+\frac{k}{\bar{a}^{2}}=\frac{8\pi G}{3}(\bar{\rho}_{v}+\bar{\rho}_{m}).
  \ee
In Einstein frame, the energy density has been rescaled and even the
energy density of matter is no more conserved.

It should be noted that the  energy measured by an observer is the
one in  Jordan-frame. Based on the rescaled metric, the relations
of the scalar factor and the Hubble parameter between the two frames
hold as
 \begin{eqnarray}
 \bar{a}=\sqrt{f}a,\,\,\,\,\bar{H}=\frac{d\bar{a}}{\bar{a}dt}=H+\frac{\dot{f}}{2f}.
 \end{eqnarray}
Then the Friedmann equation (\ref{barh2})  in the Einstein frame could
be transferred to the one in the Jordan frame
  \be
  \label{H2}
H^{2}+\frac{k}{a^{2}}=\frac{8\pi
G}{3}\left(\frac{3}{2}(\dot{B}^{2}+HB)^{2}+3V+6H(\dot{f}+H
f)+\rho_{m}\right).
 \ee
 It is just the correct Friedmann equation in the non-minimally coupled ``cosmic triad" vector field scenario.

In a short summary,  the
form of the entropy in the non-minimally coupled ``cosmic triad"
scenario is needed to directly get the Friedmann equation. Unfortunately,  such entropy is unknown. Therefore, we
transfer the  Jordan frame to the Einstein frame where the definition
of the entropy is clear. In the Einstein frame, we have obtained the
dynamical equation with the rescaled variables. At last, by
transferring these variables back to the Jordan frame, one has the
correct Friedmann equation.

The equilibrium thermodynamics could be held in  Einstein frame. As
 the exact physics in the Jordan frame is unknown, there are clearly at least two possibilities for this theory. If the
thermodynamics in Jordan frame is in equilibrium, the
thermal process are both equilibrious before and after the conformal
transformation. But, if it is a non-thermal equilibrium process in the
Jordan frame which is contrast to the Einstein frame case, the
 derived Friedmann equation is just a coincidence.
 This problem could be left to  quantum gravity.

 \section{The Generalized Misner-Sharp Energy}\label{sec5}
 Due to the strong equivalence principle,  the energy-momentum pseudo-tensor
 of gravitational field will vanish  at any point of spacetime in a locally flat coordinate.
 Therefore,  a local energy density of gravitational field does not make any sense \cite{Szabados}.
 However, there exist two well-known definitions of total energy: the Bondi-Sachs (BS)
 energy \cite{Bondi} and the  Arnowitt-Deser-Misner (ADM) energy \cite{ADM}. And,
 considering a boundary of a given region in spacetime, it is possible to define quasi-local energy, for instance, Brown-York energy \cite{York}, Misner-Sharp energy \cite{Misner:1964je}, etc. In particularly,
  at null and spatial infinity, the Misner-Sharp mass reduce to the BS and ADM energies \cite{Hayward2,Hayward3}.  When
the notion of  generalized Misner-Sharp energy (or mass) is introduced, it seems
clear to write and interpret the unified first law \cite{Hayward2,Hayward3}.

Based on
the method developed in Ref.\cite{Cai:2009qf} where the Einstein
equations are used, we will calculate the corresponding Misner-Sharp
energy $E_{M}$ which  is defined in a spherically symmetric
spacetime of the ``cosmic triad" vector field model. The integral method which is introduced in
Ref.\cite{Cai:2009qf} shows that the definition of the generalized
Misner-Sharp energy depends on a constraint condition.
 For convenience,  another form of double-null metric  is  considered in Ref.\cite{Cai:2009qf},
\begin{equation}
\label{metric3} ds^{2}=-dt^{2}+e^{2\psi (t,\rho )}d\rho
^{2}+r^{2}(t,\rho )(d\theta ^{2}+\sin ^{2}\theta d\phi ^{2}).
\end{equation}
where  $r(t,\rho )\equiv a(t)\rho $ and  $e^{\psi (t,\rho
)}=a(t)/\sqrt{1-k\rho ^{2}}$. Following the integral method, we try
to list the generalized Misner-Sharp masses.

\subsection{Minimal Coupling case}

Under the double-null metric (\ref{metric3}), the generalized
Misner-Sharp energy acts as the boundary of a finite region under
consideration in the Einstein gravity. Here, we choose the method
developed in Ref.\cite{Cai:2009qf} to calculate the
generalized Misner-Sharp mass which could be used both for the minimal
and for the non-minimal coupling cases. Based on the definition, the
general Misner-Sharp mass is
 \be
 \label{E11}
E_{M}=\frac{r}{2G}(1-h^{ab}\partial_{a}r\partial_{b}r)=\frac{r^{3}}{2G}(H^{2}+\frac{k}{a^{2}}).
 \ee
In the small-sphere limit, the leading term of $E_{M}$ is the production
of the volume and the energy density of  matter \cite{Hayward2},
 \be
 \label{E12}
  E_{M}=\rho^{(m)}V=\frac{4\pi r^{3}}{3}\left(\frac{3}{2}(\dot{B}+H
  B)^{2}+3V(B^{2})+\rho_{m}\right).
  \ee
Matching the above equations (\ref{E11}) and (\ref{E12}), the
Friedman Equation could be gotten. However, Einstein equation is used
in the derivation  of Eq.(\ref{E12}).
Therefore, it is not a surprise to get the Friedmann equation. This
calculation  demonstrates that the Misner-sharp is a consistent variable
in Einstein equation. Therefore, for the
unified first law,  the Misner-Sharp energy is also a  consistent
quantity.

\subsection{Non-Minimal Coupling case in Jordan Frame}
The generalized  Misner-Sharp mass is related to the Einstein
equation closely. And, the integral method could be used both in
Jordan  and in Einstein frames. Therefore, even in the non-minimal coupling case, we could get the generalized
Misner-Sharp mass. For  metric (\ref{metric3}), by using the
action (\ref{actionnon}), the component of the matter part of the
stress-energy tensor is
\begin{eqnarray}
&&8\pi G T_{tt}^{(m)}=3f(\frac{k}{a^{2}+H^{2}})+3H\dot{f},\,\,\,\,\,\,8\pi G T_{t\rho}^{(m)}=0\\
&& 8\pi G
T_{\rho\rho}^{(m)}=\frac{a^{2}}{1-k\rho^{2}}\left(-f(\frac{k}{a^{2}}+H^{2}+\frac{2\ddot{a}}{a})-\ddot{f}-2H\dot{f}\right)
\end{eqnarray}
and  based on the unified first law, the generalized Misner-Sharp
mass is
 \begin{equation}
  dE_{nM}=A_{s}\Psi  +WdV=C(t,\rho)dt+D(t,\rho )d\rho,
 \end{equation}
 where
 \begin{eqnarray}
  &&C(t,\rho)=4\pi r^{2}e^{-2\psi}(T_{t\rho}^{(m)}r_{,\rho}-T_{\rho\rho}^{(m)}r_{,t})=\frac{1}{2}Hr^{3}\left[f(\frac{k}{a^{2}}+H^{2}+\frac{2\ddot{a}}{a})+\ddot{f}+2H\dot{f}\right],\\
  &&D(t,\rho)=4\pi r^{2}(T_{tt}^{(m)}r_{\rho}-T_{t\rho}^{(m)}r_{,t})=\frac{1}{2}\rho^{2}a^{3}\left[3f(\frac{k}{a^{2}})+3H\dot{f}\right].
  \end{eqnarray}
Then, the energy could be calculated as:
 \be E_{nM} =\int D(t,\rho
)d\rho +\int\Big[C(t,\rho )-\frac{\partial}{\partial t} \int
D(t,\rho )d\rho \Big]dt
 \ee
If the parameters $C$ and $D$  satisfy the constraint condition
 \be
\label{intergral}
\frac{\partial C(t,\rho )}{\partial \rho
}-\frac{\partial D(t,\rho )}{
\partial t}=0,
 \ee
   the generalized Misner-Sharp mass will be gotten
  \be
  \label{dee1}
 E_{nM}=\frac{r^{3}}{2G}\left(f(B^{2})(\frac{k}{a^{2}}
 +H^{2})+H \dot{f}(B^{2})\right).
  \ee
And in the small-sphere limit of  the non-minimal coupling case, the
leading term in $E_{nM}$ is the production of volume and the energy
density of the matter
  \be
  \label{21}
 E_{nM}=\frac{4\pi r^{3}}{3}\rho^{(m)}=\frac{4\pi r^{3}}{3}\left(\rho_{m}+\frac{3}{2}(\dot{B}+H
 B)^{2}+3V\right).
  \ee
It is a
consistent result that the Friedmann equation is given out by
combining Eqs.(\ref{dee1}) and (\ref{21}).

\subsection{Non-Minimal Coupling case in Einstein Frame}

In Einstein frame, the definition of the Misner-Sharp energy gives
out the geometric representation
 \begin{eqnarray}
 \label{31}
\bar{ E}_{nM}=
\frac{r}{2G}(1-\bar{h}^{ab}\partial_{a}r\partial_{b}r)=\frac{r^{3}}{2G}\left(\frac{k}{a^{2}}
 +\bar{H}^{2}\right).
  \end{eqnarray}
And, in the non-minimal coupling case,  the total
matter contains  the redefined vector  field and the
pressureless matter. In  small-sphere limit, by using the Einstein
equation, the leading term of $\bar{E}_{nM}$ is  the production of the volume and
the energy density of  total matter
  \be
  \label{32}
  \bar{E}_{nM}=\frac{4\pi r^{3}}{3}\bar{\rho}^{(m)}=\frac{4\pi
  r^{3}}{3}\left(\bar{\rho}_{m}+\bar{\rho}_{v}\right).
  \ee
Then, combined with Eq.(\ref{31}) and (\ref{32}),    the
 Friedmann Equation (\ref{H2}) is gotten in the Einstein frame. After another conformal
transformation, we could get the correct Friedmann equation (\ref{H2}) in  Jordan
frame. The correctness of Friedmann equation makes sure that our
 arguments on the unified first law are consistent.

  Compared with Eqs.(\ref{21}) and
(\ref{32}), the generalized  Misner-Sharp energy is being rescaled.
However, the Misner-Sharp energy is corresponding to the production of the
volume and the energy density of the matter( $\rho^{(m)}V$ in Jordan
frame and $\bar{\rho}^{(m)}V$ in Einstein  frames).  The conformal
transformation extracts the freedom in the non-minimally coupled ``cosmic
triad'' vector field theory, and the energy
density and the Misner-Sharp mass are both rescaled.

\section{Conclusion}\label{sec6}
Compared with  scalar fields,  the dynamics of  vector fields are
more complicated.  In this letter, we try to find out the relations
between   thermodynamics and  ``cosmic triad" vector field
 scenario.

In the minimal coupling case of  ``cosmic triad"
scenario,  considering the entropy is proportional to the
area of  horizon in Einstein gravity,  $dS=dA/4\pi G$  is used
for the Clausius relation. Additionally, with the unified first law, we  get the correct
Friedmann equation as  expected.  However, in the
non-minimally coupled ``cosmic triad" system, there is no
corresponding entropy. Because of the similarity
between  ``cosmic triad" scenario and scalar-tensor theory,
we transformed the non-minimally coupled vector
field in the Jordan frame  to  the Einstein frame.
In Einstein frame, the form of entropy
$d\bar{S}=d \bar{A}/4\pi G$ could be  used. The
Friedmann equation was gotten successfully by using the unified
first law of thermodynamics. By matching the variables in the two
frames, the Friedmann equation is demonstrated to be correct even in the Jordan frame. Furthermore,
we calculated the generalized Misner-Sharp energy  as well which is a
key variable for the derivations of  dynamical equations.
The generalized Misner-Sharp energy is the production of the volume and the energy density of
the matter and is demonstrated to be consistent with the unified first law.

In conclusion, the unified first law which
connects  gravity and thermodynamics is a useful way to get the
Friedmann equation in the ``cosmic triad" vector field scenario.  The correct Friedmann
equation is obtained by means of the
  Einstein frame and the generalized Misner-Sharp energy.


\section*{Acknowledgements}
YZ thanks Prof. Rong-gen Cai and Dr. Ya-peng Hu for useful
discussions. We thanks the valuable comments form the anonymous referee.
 This work was supported by  the Ministry of Science and Technology of
China national basic science Program (973 Project) under grant Nos.
2007CB815401 and 2010CB833004, the National Natural Science
Foundation of China  key project  under grant No. 10935013,
the National Natural Science Foundation of China under grant
Nos.11005164 and 11073005,  the Distinguished Young Scholar Grant 10825313,
CQ CSTC under grant Nos. 2009BA4050 and 2010BB0408,
and CQ MEC under grant No. KJTD201016. YZ was partially supported by China Postdoc Grant
No.20100470237.

\end{document}